\documentclass[11pt]{article}
\usepackage{epsf}
\def\1ad{\mbox{\normalsize $^1$}}
\def\2ad{\mbox{\normalsize $^2$}}
\def\3ad{\mbox{\normalsize $^3$}}
\def\4ad{\mbox{\normalsize $^4$}}
\def\5ad{\mbox{\normalsize $^5$}}
\def\6ad{\mbox{\normalsize $^6$}}
\def\7ad{\mbox{\normalsize $^7$}}
\def\8ad{\mbox{\normalsize $^8$}}

\def\beq{\begin{equation}}                     %
\def\eeq{\end{equation}}                       %
\def\bea{\begin{eqnarray}}                     
\def\eea{\end{eqnarray}}                       
\def\npb#1#2#3{{\it Nucl. Phys.} {\bf B#1} (#2) #3 }
\def\plb#1#2#3{{\it Phys. Lett.} {\bf B#1} (#2) #3 }
\def\prd#1#2#3{{\it Phys. Rev. } {\bf D#1} (#2) #3 }

\def\bb#1{{\tt hep-th/#1}}

\def\jhep#1#2#3{{\it J. High Energy Phys.} {\bf #1} (#2) #3 }

\begin{document}

\newcommand{\sheptitle}
{Long time scales and eternal black holes\footnote{Presented at Johns Hopkins 2003 and
Ahrenshoop 2003 workshops.}}
\newcommand{\shepauthora}
{{\sc
 J.L.F.~Barb\'on}
\footnote[1]{On leave from
Departamento de F\'{\i}sica
de Part\'{\i}culas da Universidade de Santiago de Compostela, Spain.}
}

\newcommand{\shepaddressa}
{\sl
Department of Physics, CERN, Theory Division \\
 CH-1211 Geneva 23, Switzerland \\
{\tt barbon@mail.cern.ch}}

\newcommand{\shepauthorb}
{\sc
E.~Rabinovici}

\newcommand{\shepaddressb}
{\sl
Racah Institute of Physics, The Hebrew University \\ Jerusalem 91904, Israel
 \\
{\tt eliezer@vms.huji.ac.il}}
\newcommand{\shepabstract}
{We discuss the various scales determining the temporal behaviour of correlation
functions in the presence of eternal black holes. We point out the origins of
the
failure of the semiclassical gravity approximation to respect a unitarity-based
bound suggested by Maldacena. We find that the presence of a subleading
(in the
large-$N$ approximation involved) master field does restore the compliance with
one bound but additional configurations are needed to explain the more detailed
expected time dependence of the Poincar\'e recurrences and their magnitude.
}

\begin{titlepage}
\begin{flushright}
{CERN-TH/2004-059\\
{\tt hep-th/0403268}}

\end{flushright}
\vspace{0.5in}
\begin{center}
{\large{\bf \sheptitle}}
\bigskip\bigskip \\ \shepauthora \\ \mbox{} \\ {\it \shepaddressa} \\
\vspace{0.2in}
\bigskip\bigskip  \shepauthorb \\ \mbox{} \\ {\it \shepaddressb} \\
\vspace{0.2in}

{\bf Abstract} \bigskip \end{center} \setcounter{page}{0}
 \shepabstract
\vspace{0.5in}
\begin{flushleft}
CERN-TH/2004-059\\
\today
\end{flushleft}


\end{titlepage}

\newpage


\section{Introduction}

\noindent

\setcounter{equation}{0}

Hawking's semiclassical analysis of black hole evaporation suggests
that most of the information contained in initial scattering states
is shielded behind the event horizon, never to return
back to the asymptotic
region far from the evaporating black hole \cite{rsdollar}.
 In this picture, the singularity
is capable of absorbing all the infalling
 information, which is then destroyed or
transmitted to other geometrical realms, depending on one's hypotheses
about the microphysics of the singularity. From the point of view of
measurements on the Hawking radiation, the evaporation is not described
by a unitary S-matrix. Rather, quantum coherence is violated and the linear
evolution in Hilbert space takes pure states into mixed states. Still,
probability is conserved, since density matrices $\rho$ remain
Hermitian, $\rho^\dagger = \rho$, positive, $\rho >0$ and normalized,
${\rm Tr} \rho =1$ under time evolution.

The AdS/CFT correspondence \cite{rads} 
is not consistent with this picture. In this construction,
 quantum gravity
in a $(d+1)$-dimensional  asymptotically Anti-de Sitter spacetime
 (AdS) of curvature radius $R$ is {\it defined}
in terms
of  a conformal field theory (CFT) on a spatial sphere ${\bf S}^{d-1}$ of
radius $R$.
The effective  expansion parameter in the gravity side $1/N^2 \sim G_{\rm N}/R^{d-1}$, maps to
an appropriate large $N$ limit of the CFT. For example, for two-dimensional CFT's
$N^2$ is the central charge. When the
 CFT is a
gauge theory, the AdS side is a string theory, $N$
is the rank of the gauge group,  and the string perturbative  expansion in powers
of $g_s \sim 1/N$
is identified with 't Hooft's $1/N$ expansion in the gauge theory side.

According to this definition, the formation and evaporation of small black holes
with Schwarschild radius  $R_S \ll R$,
 should be described by a unitary process in terms of
the CFT Hamiltonian. Thus, there is no room for violations of coherence as a
matter of principle.
 Unfortunately, the CFT states corresponding to small black holes
are hard to describe, and it remains a challenge to put the finger on the precise
error in Hawking's semiclassical analysis in that case.

For large AdS black holes with Schwarschild radius $R_S \gg R$
one may attempt to rise to the challenge, since they 
are thermodynamically stable and can exist in equilibrium
at fixed  (high) temperatures $ 1/\beta \gg 1/R$.
Indeed, the corresponding Bekenstein--Hawking entropy
scales like that of $N^2$ conformal degrees of freedom at high energy,
\beq\label{dens} 
S \sim \sqrt{N} \;(E\,R)^{d-1 \over d} \sim N^2 \, (R/\beta)^{d-1} \;.
\eeq
Therefore, large AdS black holes with inverse Hawking temperature
$\beta \ll R$ describe
the leading approximation to the thermodynamical functions of the canonical CFT state
\beq\label{cano} 
\rho_\beta =
 {e^{-\beta H} \over Z(\beta)}\;, \qquad Z(\beta) = {\rm Tr} \,\exp(-\beta H)\;.
\eeq

\begin{figure}
\begin{center}
\epsfysize=2in
\epsffile{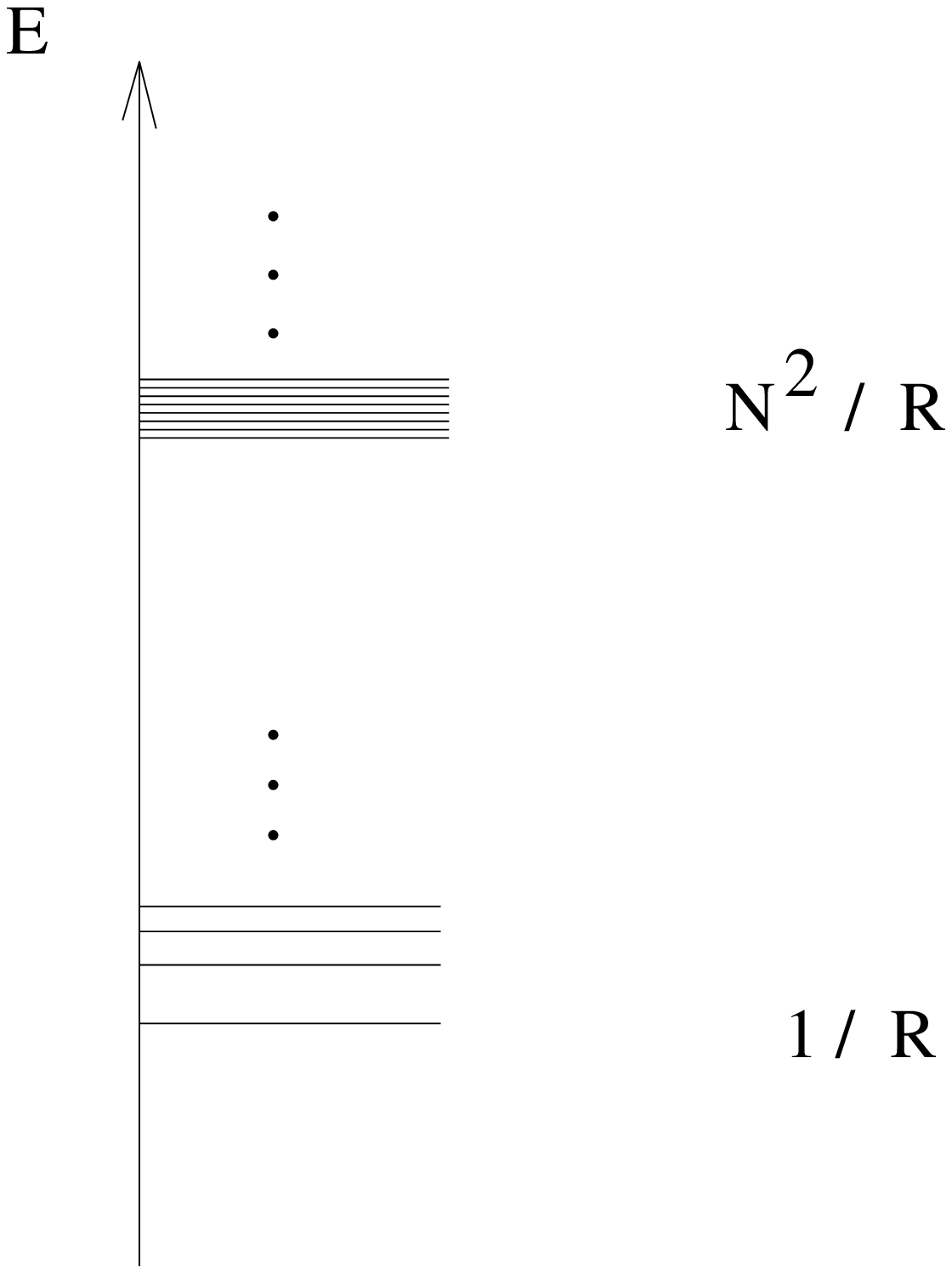}
\end{center}
\caption{
\small
\sl 
 The energy spectrum of a CFT representing ${\rm AdS}_{d+1}$ quantum gravity. The
spectrum is discrete on a sphere of radius $R$, with gap of order $1/R$.
 The asymptotic energy band of
very dense ``black hole" states
  sets in beyond energies of order $N^2 /R$. The corresponding density of states is
that of a conformal fixed point in $d$ spacetime dimensions.}
\end{figure}

This suggests that we can test the semiclassical unitarity argument by careful analysis
of slight departures from thermal 
equilibrium, rather than studying a complete evaporation instability
in the vacuum.
Ref. \cite{rmaldas} proposes to look at the very long time structure of
 correlators of the form
\beq\label{timec} 
G (t) = {\rm Tr}  \,\left[\,\rho\,A(t)\,A(0)\,\right]\;,
\eeq
for appropriate Hermitian operators $A$. In the semiclassical approximation, one expects
these correlators to decay as $\exp(-\Gamma \,t)$
 with $\Gamma \sim \beta^{-1}$. However,
because the CFT lives in finite volume,
 the spectrum is actually discrete (c.f. Fig 1), and the
correlator must show nontrivial long time structure in the form of Poincar\'e recurrences
 (see \cite{rsuscumple, rsuspodos}).
This result,
which is straightforward from the boundary theory point of view, has far reaching
consequences as far as the bulk physics is concerned.

Hence, the failure of $G(t)$ to vanish as $t\rightarrow \infty$ can be used as a criterion
for unitarity preservation
 beyond the semiclassical approximation. This argument can be made more
explicit by checking the effect of coherence loss on the long-time behaviour of $G(t)$.
Using the results of \cite{rpesk} one can simulate the required decoherence by coupling
an ordinary quantum mechanical system to a random classical noise. It is then shown in
\cite{rsus} that this random noise forces $G(t)$ to decay exponentially
for large $t$, despite having a discrete energy spectrum. This shows that the long-time
behaviour of correlators probes the strict quantum coherence of the bounded system.

At the same time, one would like to identify what kind of systematic corrections to the
leading semiclassical approximation are capable of restoring unitarity. A proposal was made
in \cite{rmaldas} in terms of topology-changing fluctuations of the AdS background. Our
purpose here is to investigate these questions and offer an explicit estimate of the
instanton effects suggested in \cite{rmaldas} (see also \cite{rsolo}).
Ultimately, this analysis should provide information about the nature of the black hole
singularity.

\section{Long-time details of thermal quasi-equilibrium}

\noindent

Poincar\'e recurrences occur in general bounded systems.
 Classically they follow from
the compactness of available phase space, plus  the preservation of
the phase-space volume in time (Liouville's theorem).
 Quantum mechanically, they follow from discreteness of the energy spectrum
(characteristic of spatially bounded systems) and unitarity, since
\beq\label{quasi} 
G_\beta (t) = {1\over Z(\beta)}
 \sum_{i,j} e^{-\beta E_i} \;|A_{ij}|^2\;e^{i(E_i -E_j)t}
\eeq
defines a quasiperiodic function of time (we have chosen the canonical
density matrix for the initial state). After initial dissipation
on a time scale $\Gamma^{-1}$, where $\Gamma$ measures the approximate width of matrix elements of
$A$ in the energy basis,
the correlator will show $O(1)$ ``resurgences" when most of the relevant phases complete
a period (c.f. Fig 2). The associated time scale is
 $t_H \equiv 1/\langle \omega \rangle$, with
$\langle \omega \rangle = \langle E_i - E_j \rangle$ an average frequency in 
(\ref{quasi}). We can estimate
$\langle \omega \rangle$ as $\Gamma /\Delta n_\Gamma$, where $\Delta n_\Gamma$ is the number of energy
levels in the relevant band of width $\Gamma$.  Introducing the microcanonical entropy
in terms of the level-number function as  $n(E)\equiv \exp S(E)$, we have
\beq\label{es} 
\Delta n_\Gamma \approx \int_{\langle E \rangle -\Gamma/2}^{\langle E\rangle +\Gamma/2} dE \,{dn\over dE}
= \int_{\langle E \rangle -\Gamma/2}^{\langle E\rangle +\Gamma/2} dE \,\beta(E)\,e^{S(E)} \approx
\Gamma \,\beta\,e^{S(\beta)}\;.
\eeq 
where we have introduced the microcanonical inverse temperature as $\beta(E) \equiv dS/dE$.

\begin{figure}
\begin{center}
\epsfysize=2.5in
\epsffile{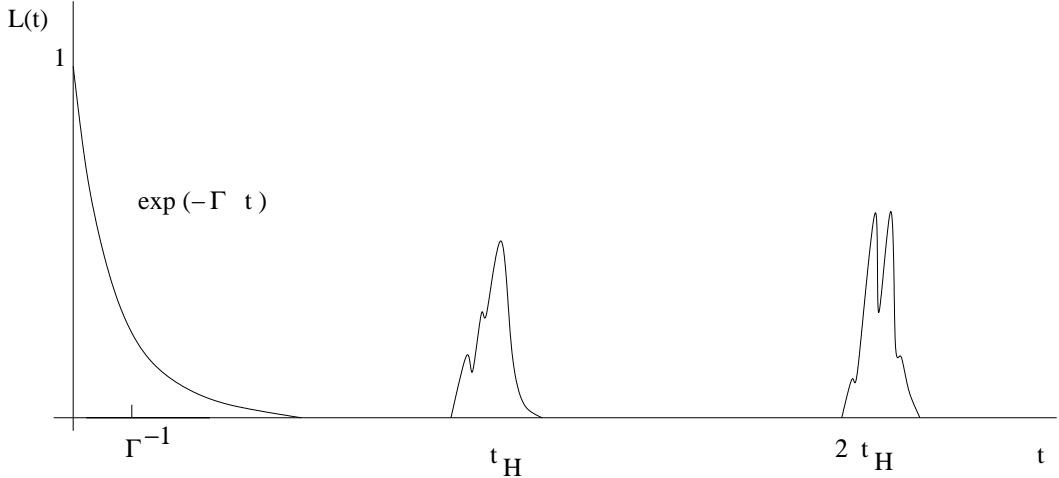}
\end{center}
\caption{
\small
\sl
 Schematic representation of
the  very long time behaviour of the normalized time correlator $L(t)$ in bounded systems.
The initial decay with lifetime of order $ \Gamma^{-1}$
is followed by O(1) ``resurgences" after the Heisenberg time $t_H \sim \beta\,\exp(S)$
has elapsed. Poincar\'e recurrence times can be defined by demanding the resurgences to
approach unity with a given {\it a priori} accuracy, and  scale like
a double exponential of the entropy.
}
\end{figure}

From this analysis we obtain an estimate
\beq\label{thes} 
t_H \sim \beta \;e^{S(\beta)}\,.
\eeq
Following \cite{rsrednicki} we call this the   Heisenberg time scale.
 Poincar\'e times
can be defined in terms of quasiperiodic returns of $G_\beta (t)$
 with a given {\it a priori} accuracy. In a
sense, the Heisenberg time is the smallest possible Poincar\'e time.

A more quantitative criterion can be used by defining a normalized positive correlator,
$L(t)$, satisfying $L(0)=1$, and its infinite time average,
\beq\label{eledef} 
L(t) \equiv \left|{G(t) \over G(0)}\right|^2, \qquad {\overline L} \equiv \lim_{T\rightarrow
\infty} {1\over T} \int_0^T dt \,L(t)\;.
\eeq 
The profile of $L(t)$ is sketched in Fig 2. The time average can be estimated by noticing
that the graph of $L(t)$ features positive ``bumps" of height $\Delta L$ and width $\Gamma$,
separated a time $t_H$, so that
\beq\label{fores} 
{\overline L} \sim {\Delta L \over  \Gamma \,t_H}\;.
\eeq 
For the case at hand $\Delta L \sim 1$,  $t_H \sim \beta\,e^S$, and we obtain
 (c.f. \cite{rsuspodos,rsus})
\beq\label{otr} 
{\overline L} \sim {e^{-S(\beta)} \over \beta \,\Gamma}\;.
\eeq 
Since both $\beta $ and $ \Gamma$ scale as  $N^0$ in the large-$N$ limit of the dual CFT, the
``recurrence index"  ${\overline L} \sim \exp(-N^2)$ scales as a nonperturbative
effect in the semiclassical approximation.

Indeed, one finds ${\overline L}=0$ in gravity
 perturbation theory in the AdS black hole background.
The reason is that the relevant eigenfrequencies $\omega$ (the so-called
normal modes of the black hole) form a continuous spectrum
to all orders in the $1/N$ expansion. 
 For a static metric
 of the form
\beq\label{bhk} 
ds^2 = -g(r)\,dt^2 + {dr^2 \over g(r)} + r^2 \,d\Omega_{d-2}^2 \;,
\eeq 
the normal  frequency spectrum follows from the diagonalization of a radial Schr\"odinger operator
 \beq\label{veun}
 \omega^2 = - {d^2 \over dr_*^2} +
 V_{\rm eff} (r_*)
   \;,
\eeq 
   with
   \beq\label{vedo} 
   V_{\rm eff} = {d-2 \over 2}\,g(r)\left({g'(r)\over r}  +
   {d-4 \over 2r^2} \,g(r) \right) + g(r)  \left(-{\nabla^2_\Omega
   \over r^2} + m^2 \right)
   \;
\eeq 
   for a scalar field of mass $m$ (analogous potentials can be deduced
for higher spin fields). Here we have defined  the
 Regge--Wheeler or   ``tortoise" coordinate
   $ dr_* = dr / g(r)
  $.

\begin{figure}
\begin{center}
\epsfysize=1.5in
\epsffile{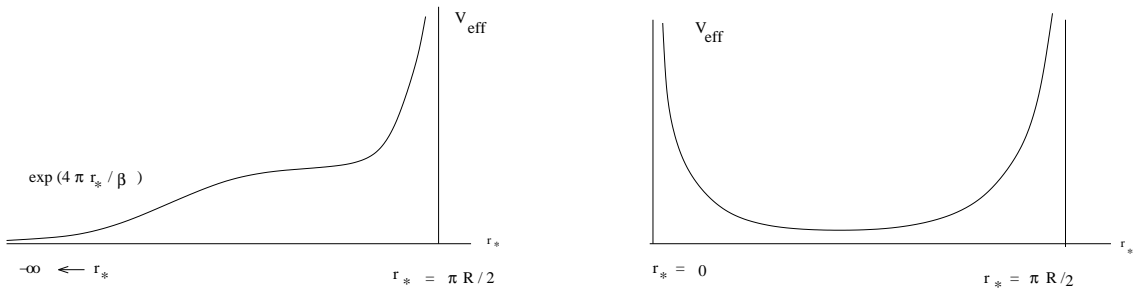}
\end{center}
\caption{
\small
\sl
 The effective potential determining the semiclassical
 normal frequency modes in a large AdS black hole
background (left).
 In Regge--Wheeler coordinates the horizon is at $r_* = -\infty$, whereas
the boundary of AdS is at $r_* = \pi R/2$ (only the region exterior to the horizon
appears). There is a universal exponential behaviour
in the near-horizon (Rindler) region. The effective one-dimensional Schr\"odinger problem
 represents a semi-infinite
barrier 
with a  continuous energy spectrum. This contrasts with
the analogous effective potential in vacuum AdS with global coordinates (right). The
domain of $r_*$ is  compact and the spectrum of normal modes is discrete with gap of
order $1/R$.}
\end{figure}

We have shown in Fig. 3 the form of the resulting effective potentials for large AdS black
holes, compared with the case of the vacuum AdS manifold.
The vacuum AdS manifold, corresponding to the choice
$g(r) = 1+r^2 /R^2$ in (\ref{bhk}),
behaves like a finite cavity, as expected.  The distinguishing feature
of the black-hole  horizon is a
 a non-degenerate
zero, $g(r_0) =0$, which induces   the universal scaling
\beq\label{univs}
 V_{\rm eff} (r_*) \;\propto \;\exp(4\pi r_* /\beta) \;\;\;\;{\rm as}\;\;\;
 r_*
\rightarrow -\infty\;, 
\eeq
 with $1/\beta
= g'(r_0) /4\pi$ the Hawking temperature and the horizon $r=r_0$ appearing
at $r_* = -\infty$.
      The
 spectrum is  discrete in pure AdS, and continuous in the AdS black hole.
Physically, this just reflects the fact that the horizon is an infinite redshift surface, so
that we can store an arbitrary number of modes with finite total energy, provided they are
sufficiently red-shifted by approaching the horizon \cite{rbrick}.
Since the thermal entropy of perturbative gravity excitations
 in the vacuum AdS spacetime scales as
$S(\beta)_{\rm AdS} \sim N^0$, we see that the perturbative Heisenberg time of
the AdS spacetime is of $O(1)$ in the large-$N$ limit, leading to ${\overline L}_{\rm AdS}
= O(1)$. On the other hand, we have ${\overline L}_{\rm bh} =0$ in this approximation.

\section{Topological diversity and unitarity}

\noindent

It is instructive to understand these perturbative results  in the
Euclidean formalism, obtained by $t=-i\tau$ in (\ref{bhk}), followed by an identification
$\tau \equiv \tau + \beta$.  The resulting metric for the vacuum AdS spacetime
has a non-contractible ${\bf S}^1$ given by the $\tau$ compact direction. We call
$Y$ this Euclidean manifold. On the other hand, the black hole spacetime with
$g(r_0)=0$ has different topology, since the thermal ${\bf S}^1$ shrinks to zero
size at $r=r_0$. The choice $1/\beta = g'(r_0)/4\pi$ ensures smoothness at
$r=r_0$. We call this Euclidean black hole manifold $X$.

The real-time correlation functions in the black hole background, $G(t)_X$,
 follow by analytic continuation from
their Euclidean counterparts. Since $X$ is a completely smooth manifold in the
$1/N$ expansion, so is the Euclidean correlator $G(it)_X$ for $t\neq 0$.
 The continuous
spectrum arising in the spectral decomposition of $G(t)_X$ is a consequence of
the contractible topology of $X$, since the Hamiltonian folliation by $\tau={\rm constant}$
surfaces is singular at $r=r_0$.

\begin{figure}
\begin{center}
\epsfysize=3in
\epsffile{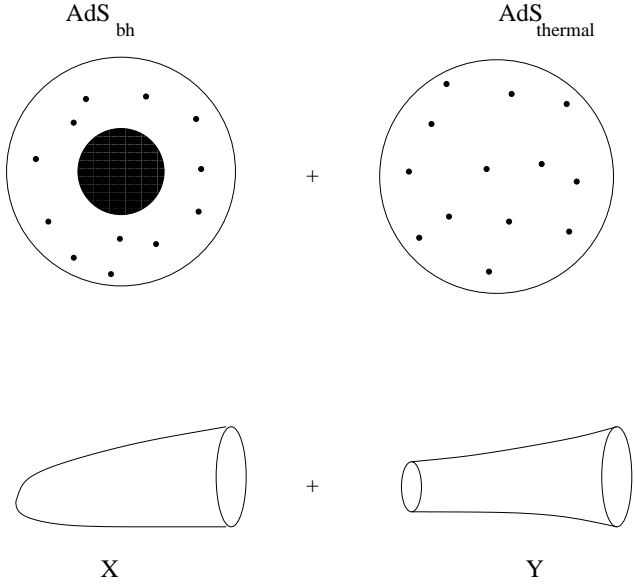}
\end{center}
\caption{
\small
\sl
 Summing over  large-scale fluctuations of the thermal ensemble in which
a black hole spontaneously turns into radiation (and viceversa) is represented in the
Euclidean formalism as the coherent sum of thermal saddle points of different topology.
The ``cigar-like" geometry $X$ represents the black-hole master field (in the CFT
language) and the cylindrical topology $Y$ represents the thermal gas of particles.}
\end{figure}

Therefore, it seems that improving on the semiclassical prediction for ${\overline L}$
requires some sort of topology-change process. The proposal of \cite{rmaldas}
  is
precisely that: instead of evaluating the semiclassical correlators on $X$, one
 should
sum coherently the contribution of $X$ and $Y$. Normally one neglects the contribution of
$Y$ on a quantitative basis (at high temperatures $R\gg \beta$).
 However, here the contribution of $X$ to ${\overline L}$
vanishes and one is forced to consider the first correction.  Since $Y$ has a discrete spectrum
in perturbation theory, the net result for ${\overline L}$ should be non-vanishing in
this approximation.
Physically, this superposition of Euclidean saddle points (or master fields, in the
language of the CFT) corresponds to large-scale fluctuations in which the AdS
black hole is converted into a graviton gas at the same temperature and viceversa.

A more detailed estimate of this ``instanton" approximation to ${\overline L}$ yields
(c.f. \cite{rsus})
\beq\label{isntl} 
{\overline L}_{\rm inst} \approx C\;e^{-2\Delta I}\;,
\eeq 
where $C= O(N^0)$, $\Delta I = I_Y - I_X$ and $I=-\log \,Z(\beta)$, calculated in the classical
gravity approximation. Since $I_Y \sim -N^0$ and $I_X \sim -N^2$, the exponential
suppression factor is of order $\exp(-2 |I_X|) \sim \exp(-N^2)$, reproducing the
expected scaling (\ref{otr}), at least in order of magnitude (however,
in general $S_X \neq -2|I_X|$, even
at high temperature).

However, the apparent success of (\ref{isntl}) turns out to be
somewhat coincidental.  If we consider the full time profile of $L(t)$ rather than
the infinite time average, we find
\beq\label{lti} 
L(t)_{\rm inst} \approx L(t)_X + C\;e^{-2\Delta I} \;L(t)_Y\;.
\eeq 
The resulting structure is shown in Figs. 5 and 6. The instanton approximation to the
normalized correlator features the normal dissipation with lifetime $\Gamma^{-1} \sim
\beta$ coming from the contribution of $X$. However, the resurgences are controlled by
$L(t)_Y$,
 damped by a factor $\exp(-2\Delta I) \sim \exp(-N^2)$, and separated a time
$t_H (Y) \sim N^0$.

\begin{figure}
\begin{center}
\epsfysize=2.6in
\epsffile{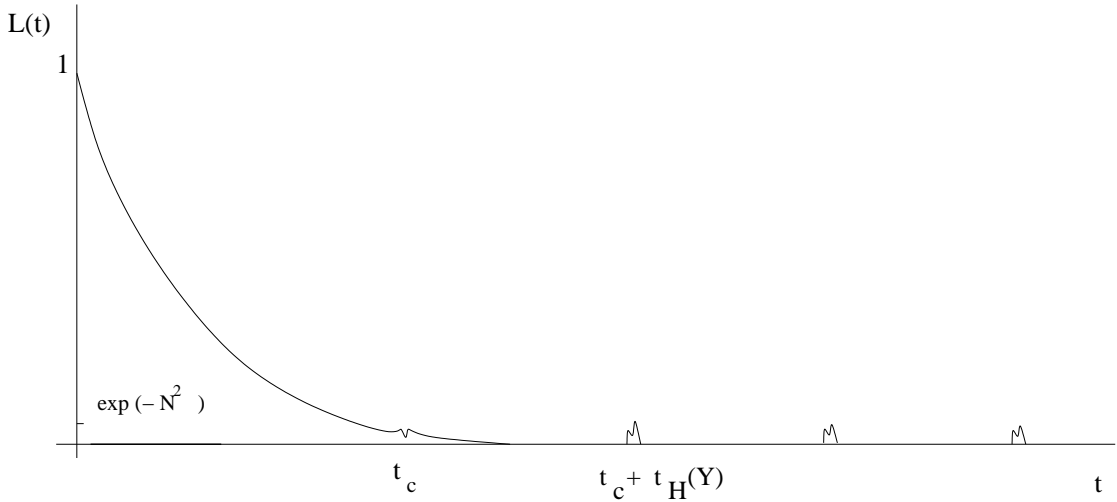}
\end{center}
\caption{
\small
\sl
The instanton approximation to the correlator $L(t)_{\rm inst}$ features
the expected exponential decay $\exp(-\Gamma \,t)$ induced by the contribution of the
$X$-manifold, whereas the resurgences
are entirely due to the  interference with the $Y$-manifold, leading to
small bumps of order $\exp(-2\Delta I) \sim \exp(-N^2)$, separated a time
$t_H (Y) \sim N^0$. These bumps are noticeable against the background of the $X$-manifold
after a time $t_c \sim \Delta I /\Gamma$.
}
\end{figure}

  Hence, the very long time
behaviour as shown in Fig. 6 is very different from the expected one, although the
infinite time average comes out  right in order of magnitude:
\beq\label{degg} 
{\overline L} \sim {\Delta L \over \Gamma \,t_H} \sim {e^{-N^2} \over \Gamma \cdot \beta}
\sim {1 \over  \Gamma \cdot \beta \,e^{N^2}}\;.
\eeq

We can also find the time scale $t_c$ for which the large-scale instantons considered
here are quantitatively important on the graph of $L(t)$. This is shown in Fig. 5 and
yields $t_c \sim \Delta I /\Gamma \sim N^2$.

\section{Conclusions}

\noindent

The study of  very long time features of correlators in black hole backgrounds is
a potentially important approach towards unraveling the mysteries of black hole
evaporation and the associated physics at the spacelike singularity.
We have seen that large scale topology-changing fluctuations proposed in
 \cite{rmaldas} 
begin to restore some of the fine structure required by unitarity, but fall short at
the quantitative level. Presumably the appropriate instantons occur on microscopic
 scales and
involve stringy dynamics.

\begin{figure}
\begin{center}
\epsfysize=2.6in
\epsffile{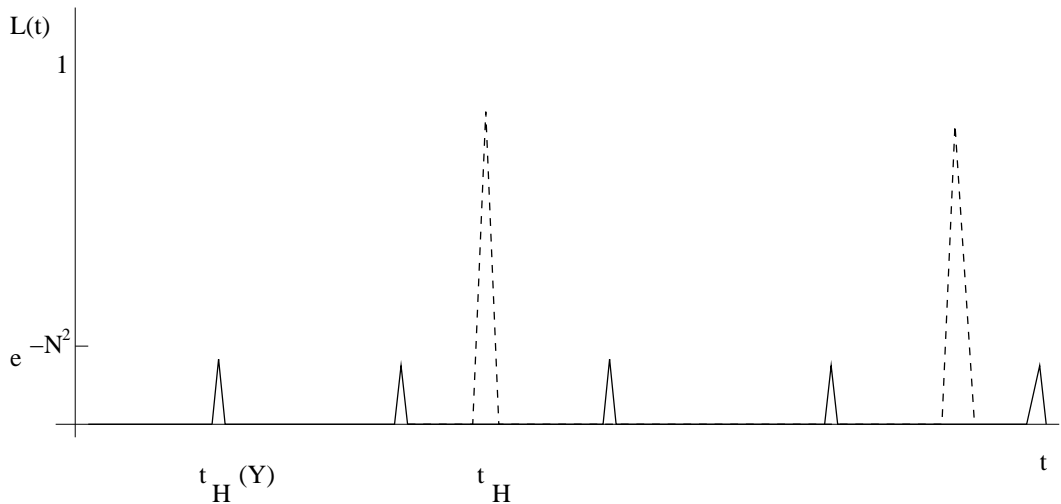}
\end{center}
\caption{
\small
\sl
Schematic representation of
the  very long time behaviour of $L(t)_{\rm inst}$ (dark line)
compared to the expected pattern for the exact quantity $L(t)$. The
resurgences of $L(t)_{\rm inst}$ occur with periods of
order $t_H (Y) = O(N^0)$ and have amplitude of order $\exp(-N^2)
\ll 1$. The expectations for the exact CFT, in the
dashed line, are  $O(1)$ resurgences with a much larger period
$t_H  \sim \exp(N^2) \gg t_H (Y)$, corresponding to tiny energy spacings of
order $\exp(-N^2)$.
 Despite the gross difference
of both profiles, the infinite time average is $O(e^{-N^2})$
for both of them.
}
\end{figure}

 While  semiclassical black holes do
 faithfully reproduce
``coarse grained" inclusive properties of the system such as the entropy
(c.f. \cite{rgh}),  additional dynamical
features of the horizon may be necessary to resolve finer details of the
information
loss problem.
Roughly, one needs a systematic set of corrections that could
generate a ``stretched horizon" of Planckian thickness \cite{rstre}.
 The crudest model of such
stretched horizon is the brick-wall model of 't Hooft \cite{rbrick}. In this phenomenological
description  one replaces the horizon by a reflecting boundary condition at Planck distance
$\epsilon \sim \ell_P$ 
from the horizon. This defines a ``mutilated" $X_\epsilon$ manifold, of cylindrical
topology, leading to a discrete spectrum of the right spacing in order of magnitude.

We have also seen that the characteristic time for large topological fluctuations to be
important is $t_c \sim O(N^2)$ in the semiclassical approximation. In 
\cite{rshenk}  it
was argued that semiclassical two-point functions  probe  the black hole singularity
on much shorter characteristic times, thereby justifying the analysis on the single standard
black hole manifold.  However, we have seen that detailed
 unitarity is only  restored on
time scales of order $t_H \sim \exp(N^2)$. Thus $t_c \ll t_H$ and we conclude that such
semiclassical analysis of the singularity is bound to be incomplete, as it misses
 whatever
microphysics is responsible for the detailed
 unitarity restoration in the quantum mechanical
time evolution.

{\bf Acknowledgement} E. R. would like to thank the KITP at Santa Barbara
for hospitality during the completion of this work, under grant of the  National Science
Foundation  No. PHY99-07949.
 The work of J.L.F.B. was partially supported by MCyT
 and FEDER under grant
BFM2002-03881 and
 the European RTN network
 HPRN-CT-2002-00325. The work of E.R. is supported in part by the
Miller Foundation, the
BSF-American Israeli Bi-National Science Foundation, The Israel Science
Foundation-Centers of Excellence Program, The German-Israel Bi-National
Science Foundation and the European RTN network HPRN-CT-2000-00122.


\end{document}